# AlphaMat: A Material Informatics Hub Connecting Data, Features, Models and Applications

Zhilong Wang[1,2†], Junfei Cai[1,2†], An Chen[1,2†], Yanqiang Han[1,2†], Kehao Tao[1,2], Simin Ye[1,2], Shiwei Wang[1,2], Imran Ali[1,2], and Jinjin Li[1,2*]

[1]National Key Laboratory of Science and Technology on Micro/Nano Fabrication, Shanghai Jiao Tong University, Shanghai, 200240, China

[2]Key Laboratory of Thin Film and Microfabrication of Ministry of Education, Department of Micro/Nano Electronics, Shanghai Jiao Tong University, Shanghai, 200240, China

[†]These authors contributed equally to this work
[*]Correspondence: Jinjin Li (lijinjin@sjtu.edu.cn)

**Graphical Abstract**

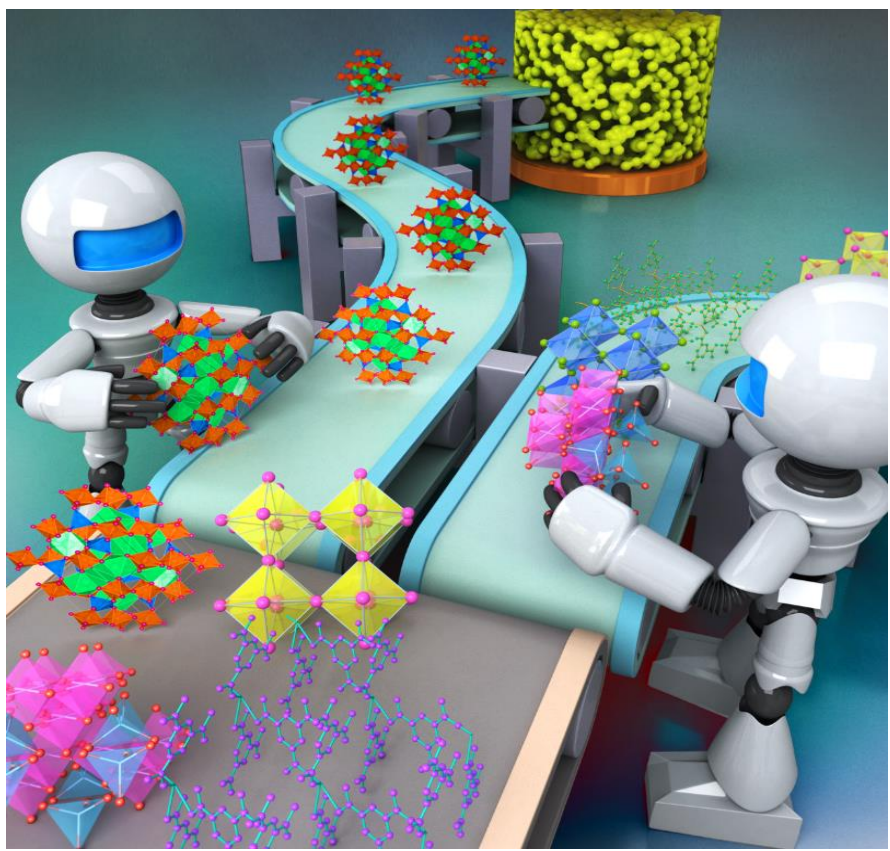

**Abstract**

The development of modern civil industry, energy and information technology is inseparable from the rapid explorations of new materials, which are hampered by months to years of painstaking attempts, resulting in only a small fraction of materials being determined in a vast chemical space. Artificial intelligence (AI)-based methods are promising to address this gap, but face many challenges such as data scarcity and inaccurate material descriptor coding. Here, we develop an AI platform, AlphaMat, that connects materials and applications. AlphaMat is not limited by the data scale (from $10^1$ to $10^6$) and can design structural and component descriptors that are effective for docking with various AI models. With prediction time of milliseconds and high accuracy, AlphaMat exhibits strong powers to model at least 12 common attributes (formation energy, band gap, ionic conductivity, magnetism, phonon property, bulk modulus, dielectric constant, adsorption energy, etc.), resulting in an unexplored material database with over 117,000 entries. We further demonstrate the ability of AlphaMat to mine and design materials, which successfully discover thousands of new materials in photonics, batteries, catalysts, and capacitors from the largest inorganic compound databases that cover all elements in periodic table. This work proposes the first material informatics hub that does not require users to have strong programming knowledge to build AI models to design materials. Users can either directly retrieve our database or easily build AI models through AlphaMat to discover and design the required materials. AlphaMat can shorten the cycle of database construction and material discovery by at least decades, and its effective use will facilitate the applications of AI technology in material science and lead scientific and technological progress to a new height.

**Introduction**

Material science has developed rapidly in the twenty-first century, both theoretically and experimentally, such as the development of gas conversion catalytic materials, the discovery of energy harvesting and storage materials, the design of information functional materials, etc[1–3]. As an interdisciplinary subject of material science and computer science, computational material science is increasingly powerful due to the significant improvement of computing power, and has become a bridge between theoretical prediction and experimental research[3–5]. Computational material science not only frees theoretical work from the bondage of analytical derivation, but also carries on the fundamental reform to the experimental research methods, which is more conducive to researchers to reveal and confirm objective laws from experimental phenomena. Currently, the modern material-simulation toolkits (*e.g.*, Vienna Ab Initio Simulation Package (VASP)[6], Quanqum Espresso[7], crystal structure analysis by particle swarm optimization (CALYPSO)[8,9], nonadiabatic molecular dynamics (Hefei-NAMD)[10], and defect and dopant ab-initio simulation package (DASP)[11], and user-friendly VASPKIT[12]) have brought computational material science to the masses in form of useful practical tools, enabling experimentalists with little or no theoretical training to perform first-principles calculations (*e.g.*, density functional theory (DFT) calculations[13,14]). Consequently, high-throughput calculations (HTC) becomes a routine approach, and accelerates the development of databases with materials (organic and inorganic crystals, single molecules and metal alloys) and properties (band gaps, formation energies, ionic conductivities, and elastical modulus[15–17]). The "Materials Genome Initiative" (MGI) proposed in 2011 pushed computaional material science into high gear[18,19], and gave birth to many material databases and platforms, such as the Materials Project (MP)[15], the Open Quantum Materials database (OQMD)[20], the Novel Materials Discovery (NOMAD)[21], and various proprietary databases from the literatures.

The establishment and sharing of these databases offer an opportunity for the emergence of the "fourth paradigm of science" and the "fourth industrial revolution", i.e. the "data-driven materials discovery"[22], the critical idea of which is the combination of big data, aritifical intelligence (AI), and material science[23–26]. The number of AI applications in material science is growing at an alarming rate, with notable success in many systems, such as batteries[27–29], solar cells[30–32], ecomaterials[33,34]. Just like the implementation of quantum mechanical (QM) computing softwares, it is necessary to develop infrastructures that combine material science and AI in order to enable both AI researchers and material experts to design materials using AI methods. Several pioneering efforts have been launched in recent years to achieve this goal[35,36]. Ward *et al.* developed an material data mining toolkit (Matminer), which offered one-stop access to multiple data sets and provided feature descritpors of components and structures for material property mining. This toolkit has become an important foundation for the joint use of AI and material data[37]. However, Matminer does not contain AI routines itself, but instead processes data format in order to make various downstream AI libraries and tools available for material science applications. In addition, using Matminer requires the basic of programming, such as Python, which is unfriendly to material designers with little programming experience. Existing material information tools have a small scope of use, weak predictive power and poor user friendliness, and therefore are not widely utilized. It is necessary to establish a material informatics platform that supports all commonly used artificial intelligence algorithms, requires no or minimal programming skill, and contains most material databases (including custom databases). In addition, the lack of data on material properties, as well as descriptors of materials, has become a challenge for materials modeling.

Here, we develop an AI platform, AlphaMat, that supports the whole life cycle of material modeling with 91 functions (data collection → data preprocessing → feature engineering → model establishment → parameter optimization → model evaluation → result analysis). AlphaMat has a higher applicability in material modeling, benefiting from component and structural descriptors. AlphaMat is the first material information platform that possesses supervised learning (SL), transfer learning (TL), and unsupervised learning (UL) simultaneously, and can tackle the tasks of material modeling without the limitation of data scale (from $10^1$ to $10^6$). In addition, AlphaMat has an interactive interface, runs locally, requires no programming experience. As typical cases, we collect and establish 12 material property databases from experiments and HTC calculations, including formation energy, metal/semiconductor, phonon property, dielectric constant, ionic conductivity, thermal conductivity, optical property, magnetism, ferroelectric property, band gap, bulk modulus, and adsorption energy (covering a total of 19,488 materials), resulting in a material database with over 117,000 entries (Supplementary Data 1–17), which can be used to enhance photoelectric conversion efficiency, improve conductivity of metallic electrode materials, promote cycle performance of batteries, discover new solid electrolyte, inhibit the shuttle effect of Li-S batteries, develop high thermal conductivity materials, solve the heat dissipation of electronic devices, etc. Compared with the time cost of experiments or calculations used to construct the database, AlphaMat saved ~46 years in learning the model and ~497 years in material discovery. The practical application in energy science demonstrates AlphaMat's ability to discover and design materials that it successfully identify 832 potential photovoltaic materials, 95 metallic electrode materails, 13 solid-state electrolytes, 58 thermal-conductivity materials, and 84 cathodes of Li-S batteries from the lagest inorganic compound database covering all elements in periodic table. By AlphaMat, users can directly search the database according to various needs; AI models can also be easily built on any data scale to discover and design materials. Following the principles of interaction, scalability, efficiency and intelligence, AlphaMat is expected to become a universal research platform to promote and accelerate the development of material science, computer science, physical and chemical science.

**Results**

**Overview and architecture**

Considering the current challenges and requirements of material modeling, AlphaMat is developed with nine core elements (Fig. 1): (1) Proprietary databases. AlphaMat aims to build database of material properties from experiments, calculations, literatures, and open databases (*e.g.*, databases of formation energy or band gap). (2) Data processing and analysis. The establishment of material data requires the unification of data format, the conversion of file format and the statistics of material properties. (3) Material descriptor design. AlphaMat can calculate suitable digitization vector or matrix to represent materials, including component and structural descriptors. (4) Quantitative structure-property relationships (QSPR). Establishment of material-property QSPR through AI models is the most important goal and pursuit of AlphaMat. (5) New materials. Based on the well-trained QSPR, new materials with suitable properties can be explored and identified. (6) Novel properties of materials that have not been reported/studied before. (7) Physical interpretability to uncover the feature importance from AI models for the design of new materials, which is the challenge and persuit of material informatics. (8) End-to-end targeted design, which is closely related to physical interpretability and establishes a pattern of input-to-output automation that facilitates practical applications. (9) Advanced applications, the ultimate goal of AlphaMat, is to promote the progress of various material systems (*e.g.*, superconducting materials, battery materials, piezoelectric materials) by discovering new materials for applications.

The organization of AlphaMat abided by the data roadmap in the research field of material informatics, from data generation to collection, learning, and application, as shown in Fig. 2. More modeling process can be found in Note S1. In AlphaMat (v0.0.6), over 90 functions have been designed (see Supplementary Information 2), and some useful tools are used (e.g., Matminer[37], Python Materials Genomics (Pymatgen)[38], Scikit-Learn[39], extreme gradient boosting decision tree (XGBoost)[40], and Mendeleev[41]). Researchers can use AlphaMat to complete the entire process of AI and material modeling. The introduction of material descriptors, AI models, and analysis tools are provided in Note S2–4.

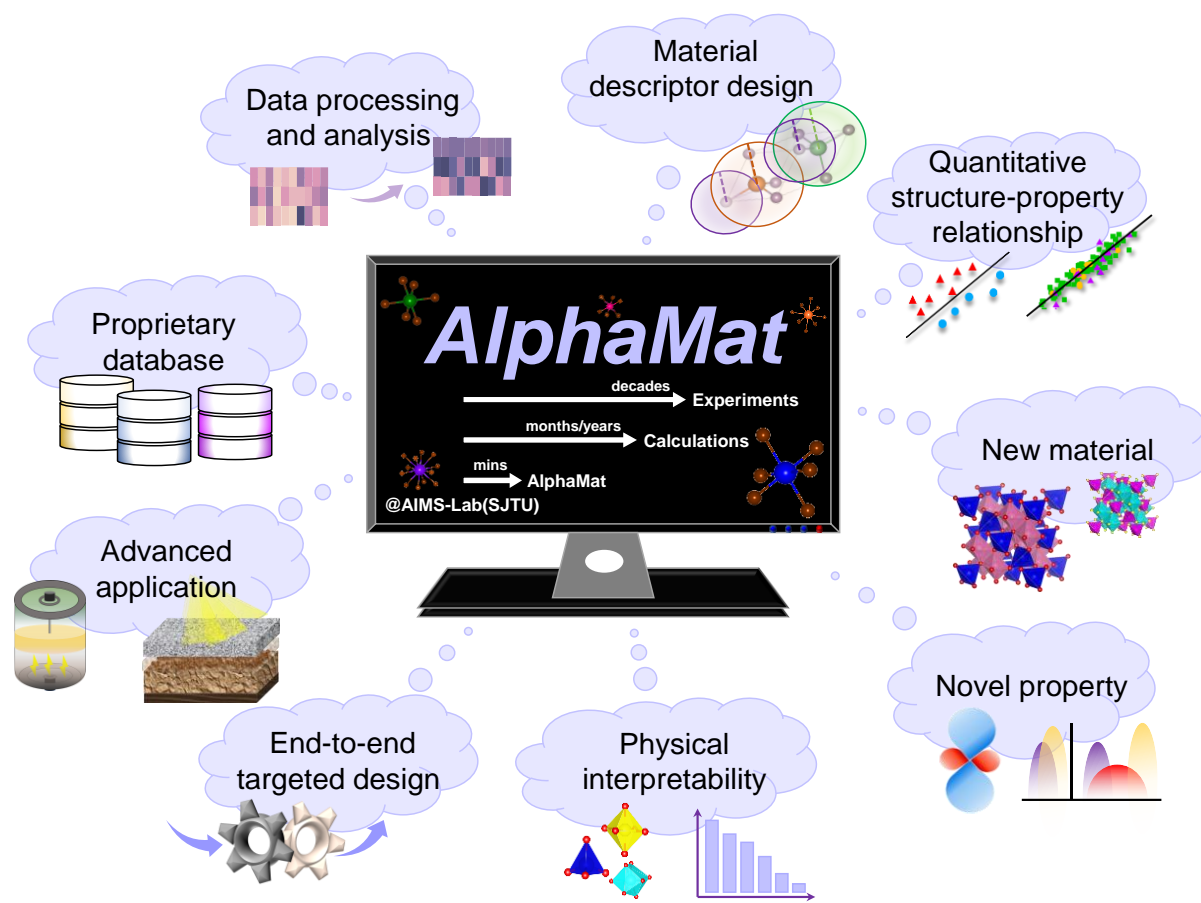

**Fig. 1 Overview of AlphaMat.** Focusing on nine elements of material informatics (proprietary databases, data processing and analysis, material descriptor design, QSPR, new materials, novel properties, physical interpretability, end-to-end targeted design, and advanced applications), AlphaMat aims to accelerate the development of materials, shorten the development cycle of materials, and reduce the cost of experimental science and traditional computations.

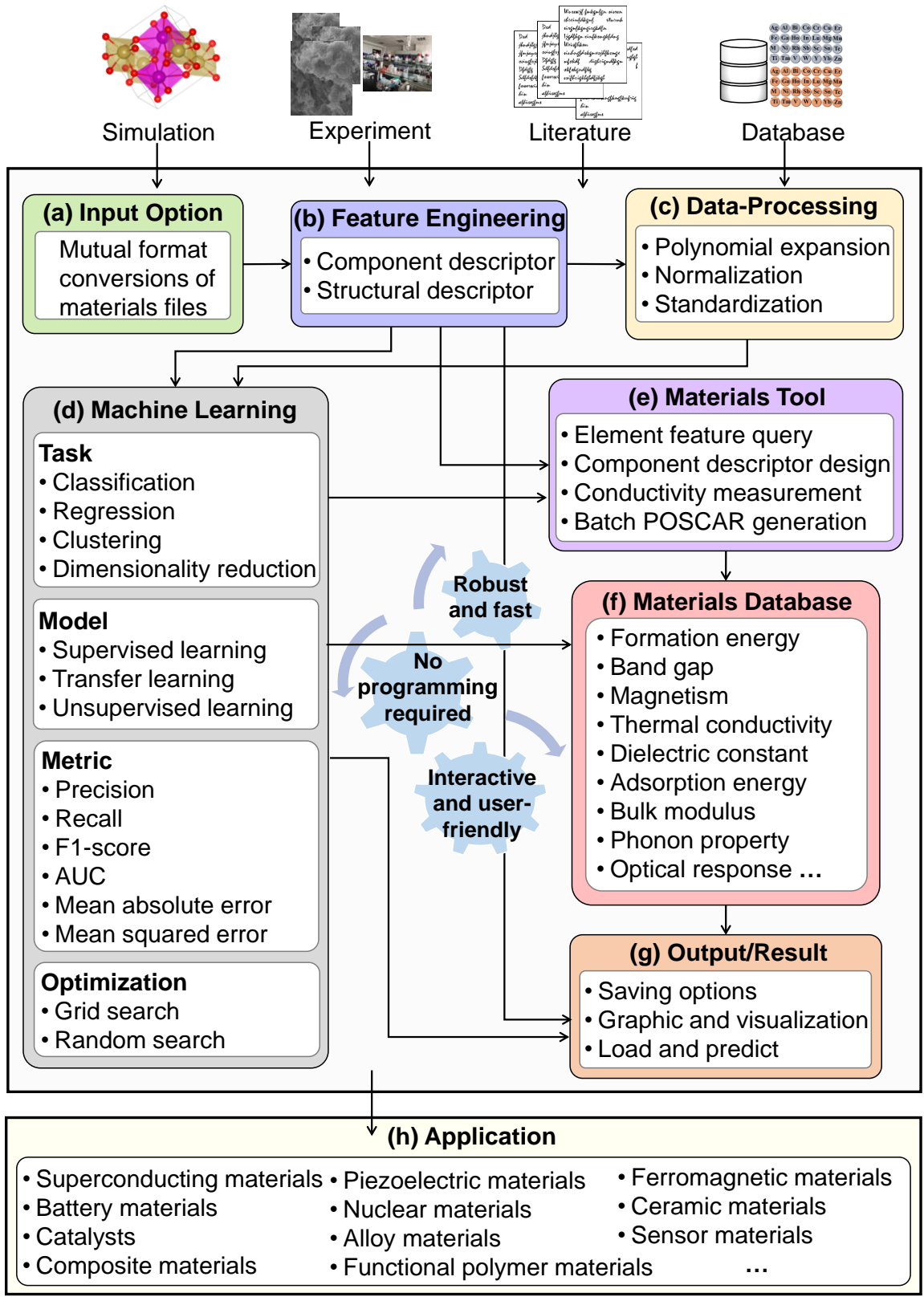

**Fig. 2 Architecture of AlphaMat. a** Input options, used to read and convert various material structure files. **b** Feature engineering, where component and structural descriptors are provided for materials representation.

**c** Data-processing, which can preprocess the obtained features. **d** Machine learning, which covers almost all current AI modeling requirements. **e** Materials tool, which integrates a variety of convenient material data processing scripts to improve efficiency. **f** Materials database, also one of the main tasks of the software. It builds proprietary databases based on different material properties and currently contains more than 50,000 data points. **g** Output/result, which can be further analyzed with various visualization tools. **h** Application, the research areas/material systems to which the whole software can be applied.

**Modeling cases**

AlphaMat provides a complete process of the data collection → data preprocessing → feature engineering → model establishment → parameter optimization → model evaluation → result analysis. Therefore, AlphaMat will play a great role in calculating material descriptors, establishing QSPR, and material screening and mining.

Here, as case studies, we use AlphaMat to predict 12 typical material properties (containing eight regression tasks and four classification tasks, see Table 1) with 19,488 data points totally (Supplementary Data 1–12), and highlight the advantages of AlphaMat in these works. The twelve material properties are formation energy ($E_f$), band gap ($E_g$), the maximum frequency of an acoustic mode at Γ (breaking of the ASR, $B_{ASR}$), dielectric constant ($\varepsilon_{poly}$), bulk modulus ($K$), ion migration activation energy ($E_a$), thermal conductivity ($\kappa$), second harmonic generation (SHG) responses, metals/semiconductors, ferroelectric/non-ferroelectric materials (Ferro/Non-ferro), strong/weak adsorption energy ($\Delta E$), and ferromagnetic/antiferromagnetic materials (FM/AFM). It is worth noting that we choose the component descriptor of element property (a 120-element vector) as the material descriptor, which is defined by Meredig *et al*. and integrated in AlphaMat (instruction of 805)[42]. XGBoost model is applied for model training, which has been widely used in materials science[40,43,44]. The descriptions of XGBoost are shown in 12104 (for classification tasks) and 12204 (for regression tasks) in Supplementary Information 2. The data is split into training set (80%) and testing set (20%). As shown in Table 1 and Fig. S1–S12, the Pearson correlation coefficients ($P_C$) of eight regression models are from 0.675 to 0.933, with an average of 0.843, and the precision of four classification models are from 0.82 to 0.93, with an average of 0.868. More details and applications are provided in Supplementary Information. These typical case studies demonstrate the strong modeling ability of AlphaMat in material property predictions and material discovery.

The 19,488 data points currently used for modeling are just the tip of the iceberg in the vast material space, and there are hundreds of millions of materials and properties to be explored. For example, in the MP database[15], the formation energies and bulk modulus of 144,595 materials data can be modeled by AlphaMat, saving approximately 11.66 years of computational cycles. In addition, the band gap prediction model established by AlphaMat can predict the band gaps of ~68,000 materials (with unknown band gaps in the MP database[15]) at the experimental level (each takes 64 h[45]), which will save around 497 years of experimental cycle. For the highly time-consuming determination of ionic conductivity (each takes ~6.76 days[44]) and adsorption energy (each takes 2.14 days[46]), AlphaMat exhibits even greater efficacy in systems with data scale > $10^5$, and saves the computational cycle of 58 years to 185 years based on conservative estimates. It can be seen that, based on the existing experimental/computational data, modeling based on AlphaMat can greatly shorten the experimental/computational cycle for new material discovery. We can foresee that AlphaMat will shine in the era of material informatics.

**Table 1. Twelve case studies.** The material features as ML inputs are calculated using component descriptors, and the classification tasks and regression tasks are implemented by training XGBoost models. The performance on testing data is selected to evaluate the AlphaMat performance. It is noted that these case studies are meant to provide macro guides for the users of AlphaMat to facilitate a complete employment of the available materials database.

| Materials property | Task | Data scale | AlphaMat cost | Exp/Cal cost | AlphaMat performance | Application | Ref |
|---|---|---|---|---|---|---|---|
| $E_f$ (eV/atom) | $R_1$ | 3,483 Cal. points | 3.92 mins | 102.51 days | $P_c$ = 0.877, MAE = 0.221, RMSE = 0.342 | All material systems | 15 |
| $E_g$ (eV) | $R_2$ | 3,895 Exp points | 4.38 mins | 28.46 years | $P_c$ = 0.933, MAE = 0.347, RMSE = 0.522 | Battery materials, catalysts, photodetectors, Spintronic devices, | 47 |
| $B_{ASR}$ (cm$^{-1}$) | $R_3$ | 1,515 Cal. points | 1.70 mins | 44.59 days | $P_c$ = 0.878, MAE = 2.278, RMSE = 4.381 | Superconducting materials, Thermoelectric materials, etc. | 48 |
| $\varepsilon_{poly}$ | $R_4$ | 1,028 Cal. points | 1.44 mins | 30.27 days | $P_c$ = 0.675, MAE = 3.265, RMSE = 4.745 | Ferroelectric Materials, Supercapacitors, etc. | 49 |
| $K$ (GPa) | $R_5$ | 373 Cal. points | 0.42 mins | 21.80 days | $P_c$ = 0.895, MAE = 15.853, RMSE = 1.248 | Battery materials, photodetectors, memory, etc. | 15 |
| $E_a$ (eV) | $R_6$ | 1,109 Cal. points | 1.25 mins | 46.21 days | $P_c$ = 0.803, MAE = 0.402, RMSE = 0.604 | Battery materials, etc. | 50 |
| $\kappa$ (W m$^{-1}$ K$^{-1}$) | $R_7$ | 128 Exp. points | 0.14 mins | 341.33 days | $P_c$ = 0.829, MAE = 0.293, RMSE = 0.465 | Thermoelectric materials, etc. | 51 |
| SHG responses (pm V$^{-1}$) | R$_8$ | 291 Cal. points | 0.33 mins | 877.26 days | $P_c$ = 0.854, MAE = 0.963, RMSE = 1.524 | Mid-infrared nonlinear optical materials, etc. | 52 |
| Metal/Semi conductor | C$_1$ | 6353 Exp. points | 7.14 mins | 46.41 years | Precision = 0.93, recall = 0.93, F1-score = 0.93, AUC = 0.93 | Battery materials, catalysts, Spintronic devices, etc. | 47 |
| Ferro/Non-ferro | C$_2$ | 1,028 Cal. points | 1.44 mins | 30.27 days | Precision = 0.87, recall = 0.87, F1-score = 0.87, AUC = 0.87 | Ferroelectric Materials, catalysts, photodetectors, etc. | 49 |
| Strong/Weak $\Delta E$ | C$_3$ | 65 Cal. points | 0.07 mins | 139.05 days | Precision = 0.85, recall = 0.85, F1-score = 0.85, AUC = 0.85 | Battery materials, catalysts, etc. | 46 |
| FM/AFM | C$_4$ | 220 Cal. points | 0.25 mins | 19.43 days | Precision = 0.82, recall = 0.80, F1-score = 0.80, AUC = 0.80 | Spintronic devices, topological quantum materials, etc. | 17 |

R: Regression tasks

C: Classification tasks

Cal: Calculated data

Exp: Experimental data

$P_C$: Pearson correlation coefficient

MAE: Mean absolute error

RMSE: Root mean squared error

AUC: Area under receiver operator characteristic (ROC) curve

**Practical applications in high-performance materials**

Twelve case studies demonstrate AlphaMat's capabilities in material modeling. Here, with different material property modeling, we present several practical examples about electrode materials, photoelectric materials, solid-state electrolyte materials, and thermal-conductivity materials, etc.

**(1) Practical applications based on $E_g$.** $E_g$ is the key characteristic of electronic materials. For example, in perovskite solar cells, the hole transport layer (HTL) and the electron transport layer (ETL) should have appropriate $E_g$ (0.9–1.6 eV) to ensure the efficient transmission of holes and electrons and the implement of optimal optical conversion efficiency[53,54]. Electrode materials generally have high conductivity, i.e., $E_g$ = 0[55,56], while solid-state electrolytes require extremely low electronic conductivity, i.e., $E_g$ > 3.5 eV[45,57,58]. Thus, accurately determining $E_g$ is the key to select and accelerate the development of materials.

MP database contains 144,595 data entries[15], among which the studies of mono-element compounds are quite mature, while the laboratory synthesis of multi-element compounds is challenging. Therefore, binary ($B_C$), ternary ($T_C$), quaternary ($Q_C$), and pentabasic ($P_C$) compounds are selected from MP to establish the initial data set. In addition, thermal stability is the most basic property of materials, so we exclude materials with convex hull energy ($E_{hull}$) greater than 0, leaving 32,858 materials in the end (5,039 $B_C$, 19,257 $T_C$, 7,287 $Q_C$, 1,275 $P_C$). Among the 12 case studies in Table 1, $C_1$ can distinguish metals ($E_g = 0$) and semiconductors ($E_g > 0$), $R_2$ can predict the $E_g$ for semiconductors. By using element property as the material descriptor (805 in AlphaMat), we make use of well-trained $C_1$ and $R_2$ models for searching new materials.

As shown in Fig. 3**a**, using t-distributed stochastic neighbor embedding (t-SNE) method, the sites are colored with their compound types, and compounds with different number of element types can be distinguished, as the sites of $P_C$, $Q_C$, $T_C$, and $B_C$ are stacked on top of each other. In MP database, the $E_g$ values of 32,858 compounds are calculated based on semi-empirical or low-precision PBE functional, which deviates greatly from the experimental value (the deviation is 1.0–2.0 eV generally) and are difficult to be directly used in the actual screening of materials[17,59]. By using band gap-based models $C_1$ and $R_2$, we can rapidly predict (or update) the $E_g$ of 32,858 compounds. Our well-trained $C_1$ has a prediction accuracy of 93% for identifying metals and semiconductors, and the $P_C$ between the $E_g$ predicted by $R_2$ and the experimental value is 0.933, and the MAE is only 0.347 eV (see Table 1). Therefore, the two models are of great significance to update and reuse materials in MP database. As shown in Fig. 3**b**, the sites are colored with their $E_g$ values predicted by $C_1$ and $R_2$, where the sites with large values (> 3.0 eV) are mainly concentrated on the right side of the t-SNE plot. This phenomenon can be associated Fig. 3**a**, as the types of compound elements increase, the new elements introduced are mainly non-metallic elements, such as O, S, F, Cl, Br, etc., leading to the weakening of the electronic conductivity of the material.

Fig. 3**c** shows the correlation of MP calculated $E_g$ and predicted $E_g$. It can be seen that the $E_g$ changes of most materials are less than 2.0 eV (blue and purple dos), and the updated $E_g$ is often larger (green, yellow, and red dots), which is consistent with the conclusion that the $E_g$ calculated by low-precision functional is seriously underestimated[17,60,61]. Then, we identify 832 materials with $E_g$ of 0.9–1.6 eV that can be used as photoelectric materials (HTLs, ETLs, photocatalysts, etc.), 13 materials containing $Li^+$ with $E_g > 3.5$ eV that are solid-state electrolyte candidates, as provided in Table S1 and S2. In addition, for searching the electrode materials, excellent electronic conductivity with $E_g = 0$ is necessary, as well as high mechanical properties. Referring to the shear modulus and bulk modulus of commercialized materials $LiNi_{0.3}Mn_{0.3}Co_{0.3}O_2$ (NMC333), $LiNi_{0.4}Mn_{0.4}Co_{0.2}O_2$ (NMC442), $LiNi_{0.5}Mn_{0.3}Co_{0.2}O_2$ (NMC532), $LiNi_{0.6}Mn_{0.2}Co_{0.2}O_2$ (NMC622), $LiNi_{0.8}Mn_{0.1}Co_{0.1}O_2$ (NMC811)[62,63], we further select 95 materials with shear modulus > 67 GPa and bulk modulus > 85 GPa as candidates for electrode materials, as provided in Table S3.

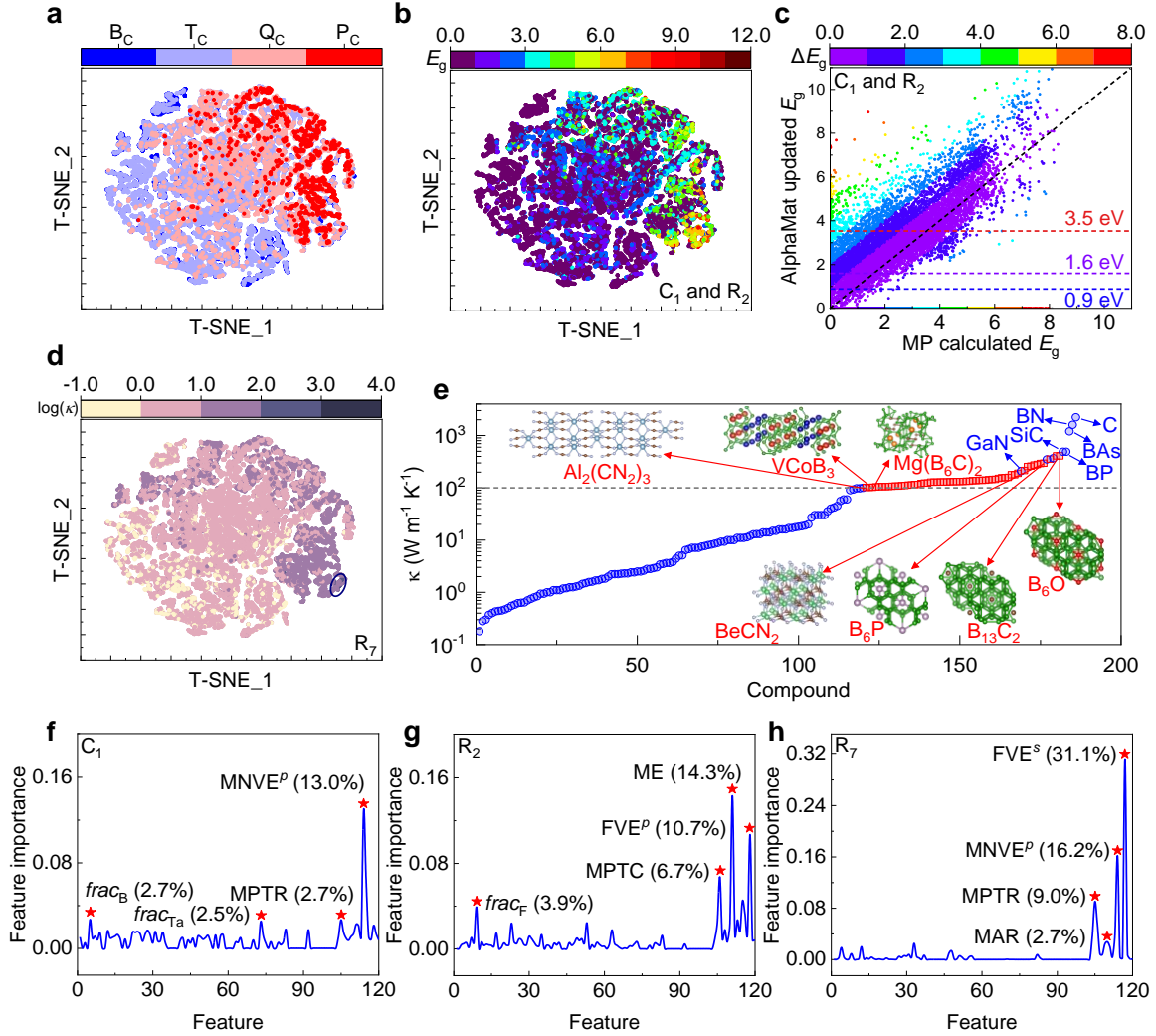

**Fig. 3 Material discovery through $C_1$, $R_2$, and $R_7$ models established by AlphaMat. a** t-SNE plot, where the sites are colored with their compound type. **b** t-SNE plot, where the sites are colored with their $E_g$. **c** Correlation of predicted $E_g$ and calculated $E_g$, the color bar denotes the $E_g$ changes. **d** t-SNE plot, where the sites are colored with their thermal conductivity, the oval mark determines the compounds with high thermal conductivities. **e** Discovered compounds with high thermal conductivities (red squares) and the known compounds (blue circles). **f** Embedded feature importance of $C_1$. **g** Embedded feature importance of $R_2$. **h** Embedded feature importance of $R_7$.

**(2) Practical applications based on $\kappa$.** $\kappa$ is an important thermal property of electronic materials and electronic devices. Materials with high $\kappa$ (e.g., C, 2235 W m$^{-1}$ K$^{-1}$; BN, 1600 W m$^{-1}$ K$^{-1}$) can be used to solve the heat dissipation problem of electronic products, and the development of new thermal conductivity materials will provide strong support for future space exploration activities and ocean exploration activities[64,65]. Accurately determining $E_g$ and $\kappa$ is critical to select and accelerate the development of new materials for practical applications. Among the 12 case studies in Table 1, $R_7$ can predict the $\kappa$ of given materials. By using element property as the material descriptor, we make use of well-trained $R_7$ models for searching materials. As shown in Fig. 3d, the t-SNE plot shows that most materials have very small $\kappa$ (< 100

W m$^{-1}$ K$^{-1}$), and materials with thermal conductivity > 100 W m$^{-1}$ K$^{-1}$ are very concentrated (see the oval mark). Fig. 3**e** shows the discovered materials (red squares) with high $\kappa$, such as B$_6$O (408.7 W m$^{-1}$ K$^{-1}$), B$_{13}$C$_2$ (407.7 W m$^{-1}$ K$^{-1}$), B$_6$P (355.0 W m$^{-1}$ K$^{-1}$), and BeCN$_2$ (296.0 W m$^{-1}$ K$^{-1}$). The new thermal-conductivity materials can be comparable to the famous GaN (210.0 W m$^{-1}$ K$^{-1}$), which has a broad prospect in the application of optoelectronics, high temperature high power devices and high frequency microwave devices.

The predicted $E_g$ and $\kappa$ of 32,858 materials at experimental level are provided in Supplementary Data 13–14, this may be of wide interest to the experimental community in multiple areas of research (batteries, catalysis, electronics, etc.). In addition to establish the high-precision QSPR, AlphaMat also provides the interpretability of the model, which is a unique feature. Fig. 3**f–h** shows the embedded feature importance of $C_1$, $R_2$, and $R_7$, respectively. For $C_1$, the mean number of valence electrons of $p$ orbitals (MNVE$^p$, 13%) in compounds and the mean of periodic table rows (MPTR, 2.7%) play a key role in distinguishing metals from semiconductor materials. This is of guiding significance for the design of corresponding materials. The fraction of B (*frac*$_B$, 2.7%) and Ta (*frac*$_{Ta}$, 2.5%) are also important due to the compounds containing B in the data are mainly semiconductors, while those containing Ta are metallic in training data set. For $R_2$, the mean electronegativity (ME, 14.3%), the fraction of valence electrons of $p$ orbitals (FVE$^p$, 10.7%), the mean of periodic table columns (MPTC, 6.7%), and the fraction of F (*frac*$_F$, 3.9%) are relatively important for predicting $E_g$ values. For $R_7$, the fraction of valence electrons of $s$ orbital (FVE$^s$, 31.1%) and MNVE$^P$ (16.2%) are particularly important for thermal conductivity prediction, which is consistent with the phenomenon that heat conduction is mainly the diffusion of free electrons from the high end to the low end, resulting in heat flow. These key features are of great significance for further directed design of functional materials.

**(3) Practical applications based on Δ*E*.** UL methods are based on unlabeled data, can completely overcome the obstacle of scarce material attributes. However, UL module is still a gap in all the existing material modeling. In above case studies, the data scale of Δ*E* between AB$_2$-type 2D materials and Li$_2$S$_6$ is few (only 65 entries), which is not conducive to establish the QSPR. The search for materials with strong adsorption (|Δ*E*| > 1.0 eV) for Li$_2$S$_6$ is helpful to discover new cathode materials for lithium-sulfur (Li-S) batteries and inhibit the "shuttle effect". Here, we demonstrate an unsupervised learning method for discovering new cathodes for Li-S batteries. Total 826 stable AB$_2$-type compounds are selected from the 2DMatPedia database, of which 65 materials have known adsorption energies with Li$_2$S$_6$, and the remaining 761 are unknown[66]. Fig. 4**a** shows the bottom-up tree diagram (dendrogram) by using agglomerative hierarchical clustering (AHC) algorithm in AlphaMat, where a suitable partition line is selected and the 826 AB$_2$-type compounds are classified into seven group (see Fig. S13, from $G_1$, $G_2$, …, to $G_7$). We map 65 known Δ*E* to the dendrogram, and compounds marked by green, orange and red are promising according to ideal thresholds (-1.0 eV)[46]. The clustering of AB$_2$-type compounds provides physical insights into understanding of compounds exhibiting proper adsorption energies for Li$_2$S$_6$. Fig. 4**b** gives the statistic of known and unknown compounds each group, $G_4$ has the most compounds of 319, while $G_3$ has the fewest compounds of 29, indicating that a targeted study of these groups would significantly narrow down the initial scope (761 unknown compounds). Fig. 4**c** shows the ratio of known compounds (black line) and the ratio of desired compounds (blue line) of each group. Notably, in $G_1$, $G_3$, and $G_5$, the ratio of desired compounds to known compounds is 100%, which is much higher than that in other groups. This phenomenon can also be observed in Fig. 4a. These suggest that the unknown compounds in $G_1$, $G_3$, and $G_5$ are worthy of further investigation (142 compounds in total), and that they may also be potential cathode materials for Li-S batteries. The violin plots of the known $\Delta E$ shown in Fig. 4d further reveal that $G_5$ is of high research value because of its higher average absolute adsorption energy value (1.62 eV). As a result, the scope of exploration narrowed from 761 compounds to 84 compounds, as provided in Fig. S14–16. More details about the position of the partition line are discussed in Note S18.

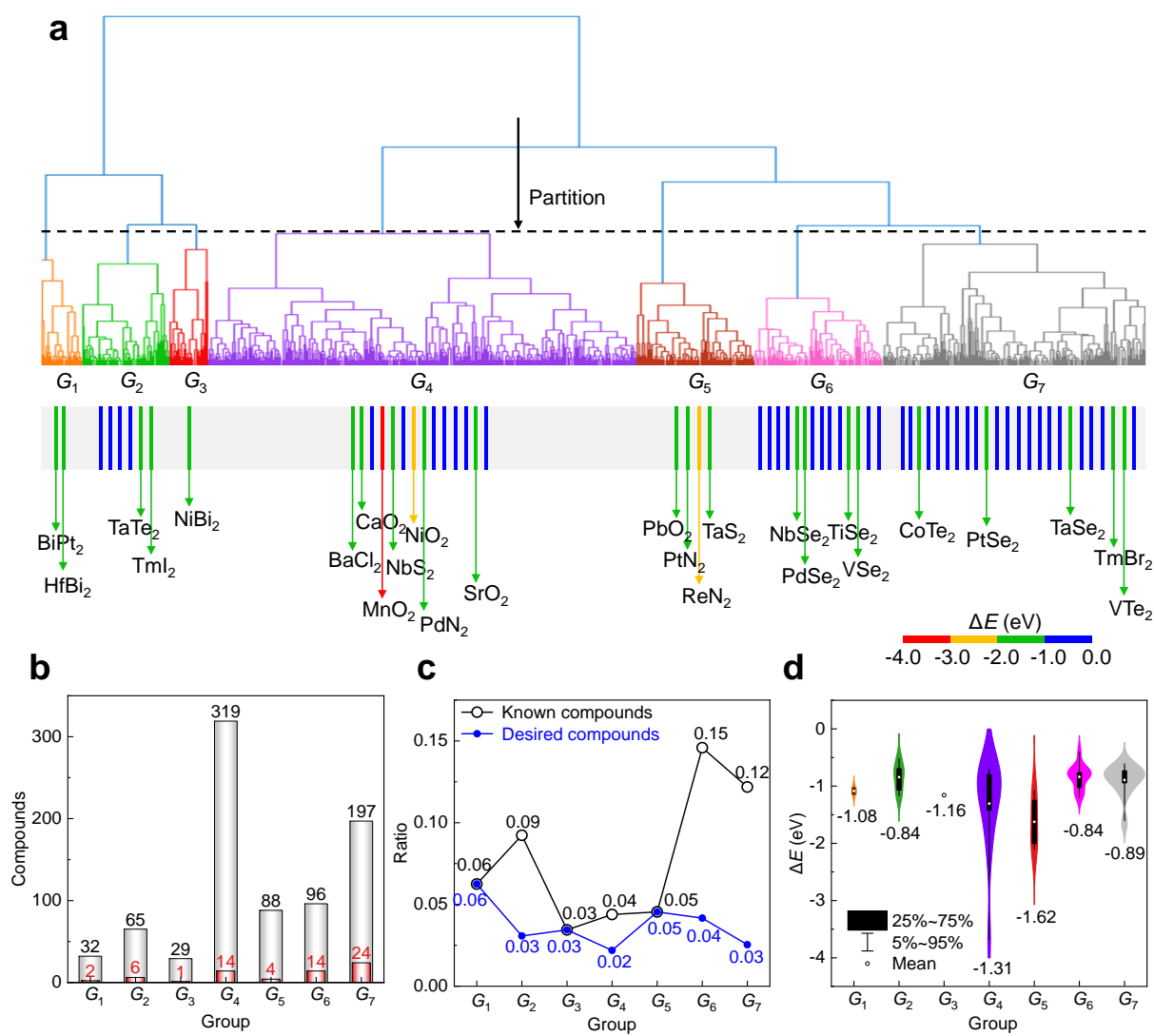

**Fig. 4 Unsupervised discovery of AB$_2$-type compounds as cathodes for Li-S batteries. a** Dendrogram generated by the AHC method in AlphaMat. The dashed line shows the position where all compounds are partitioned into seven groups, marked as $G_1$–$G_7$ from left to right and distinguished by different colors. **b** Statistic of known compounds in each group. Gray bars represent the number of compounds in each group after clustering, and red represents the number of known compounds in each group. **c** Ratio of known compounds (black line) and the ratio of desired compounds (blue line) of each group. **d** Violin plots of $\Delta E$ in seven groups. The outer shells of the violins bound all data, narrow horizontal lines bound 90% of the data, thick horizontal lines bound 50% of the data, and white dots represent means.

**Conclusions**

The challenges of material informatics prompt us to develop an advanced computational infrastructure. In this work, we present an AI paltform that supports the whole life cycle of material modeling, including data analysis, feature engineering, model establishment and optimization, evaluation to result analysis. The proposed AlphaMat integrates supervised SL, TL, and UL simultaneously, which can tackle the tasks in material science more comprehensively. Furthermore, AlphaMat establishes 17 proprietary databases with more than 117,000 material-property entries. Since AlphaMat runs locally, the training of its AI models is not limited by the scale of data sets (from $10^1$ to $10^6$). Consequently, AlphaMat will accelerate the innovative discovery of new materials, new functions, and new principles, compard to the trial-and-error experiments and high-throughput calculation methods. 12 case studies of material modelings (formation energy, band gap, magnetism, adsorption energy, thermal conductivity, and ionic conductivity, etc) demonstrate the effectiveness and usefulness of AlphaMat, and the practical application in searching high-performance materials demonstrates AlphaMat's ability to mine and design materials that it successfully identify thousands of new materials for use in various systems (photonics, batteries, catalysis, and capacitors, etc.) from the largest inorganic compound databases that covers all elements in periodic table. Using AlphaMat, users can either directly retrieve our database or easily build AI models to discover and design materials.

It should be mentioned that ML is only as good as the data it is trained on and that going beyond the training set is highly likely to fail dramatically, Therefore, the prediction results of the ML model will be uncertain to a certain extent. In the face of more complex problems, traditional computational methods or experimental methods are also needed for further verification. But at the very least, ML offers specific candidates to speed up material development.

Further, we will continue to improve and release AlphaMat to address the challenges commonly encoutered in material modeling: (1) continuously expand the databases according to the material properties to alleviate the challenge of data scracity; (2) innovate component and structural descriptors to represent the materials, improve model accuracy and make models interpretable; (3) combine frontier AI algorithms timely to imporove the performance of the models; and (4) add more convenient tools and visualization interface to improve the efficiency for processing material data. We hope that the continually released AlphaMat will deeply unite material science and AI apporaches, and become an essential tool in science researches.

**Data availability**

More details and tutorials on using AlphaMat are also available from Supplementary Information. More database construction will be incorporated in the future release of AlphaMat, please visit our website: http://www.aimslab.cn/item_25589209_2780253.html.

**Code availability**

The codes to run AlphaMat are avaliable from our website: http://www.aimslab.cn/item_25589209_2780253.html.

**Acknowledgements**

The work is supported by the National Key R&D Program of China (No. 2021YFC2100100), the National Natural Science Foundation of China (No. 21901157), the SJTU Global Strategic Partnership Fund (2020 SJTU-HUJI) and the National Key Laboratory of Science and Technology on Micro/Nano Fabrication, China. We also acknowledge the important contributions to the development of AlphaMat code from the following group members: Lin Zhang, Xirong Lin, Sicheng Wu, Zehao Yu, Jiequn Tang.

**Competing of interests**

The Software Copyrights have been obtained in China for AlphaMat with No. 2022SR0405168 and No. 2022SR1364423.

**Author contributions**

**Zhilong Wang, Junfei Cai, An Chen and Yanqiang Han:** Conceptualization, Methodology, Visualization, Data curation, Writing - original draft. **Kehao Tao:** Methodology, Visualization. **Simin Ye, Shiwei Wang, Imran Ali:** Data curation. **Jinjin Li:** Conceptualization, Methodology, Supervision, Resources, Writing - review & editing. All authors commented on the manuscript. These authors contributed equally: **Zhilong Wang, Junfei Cai, An Chen, and Yanqiang Han**.

**References**

1. Daehn, K. *et al.* Innovations to decarbonize materials industries. *Nat. Rev. Mater.* **7**, 275–294 (2021).
2. Marzari, N., Ferretti, A. & Wolverton, C. Electronic-structure methods for materials design. *Nat. Mater.* **20**, 736–749 (2021).
3. Louie, S. G., Chan, Y.-H., da Jornada, F. H., Li, Z. & Qiu, D. Y. Discovering and understanding materials through computation. *Nat. Mater.* **20**, 728–735 (2021).
4. Correa-Baena, J.-P. *et al.* Accelerating materials development via automation, machine learning, and high-performance computing. *Joule* **2**, 1410–1420 (2018).
5. Hammes-Schiffer, S. & Galli, G. Integration of theory and experiment in the modelling of heterogeneous electrocatalysis. *Nat. Energy* **6**, 700–705 (2021).
6. Kresse, G. & Furthmüller, J. Efficient iterative schemes for ab initio total-energy calculations using a plane-wave basis set. *Phys Rev B* **54**, 11169–11186 (1996).
7. Giannozzi, P. *et al.* QUANTUM ESPRESSO: a modular and open-source software project for quantum simulations of materials. *J. Phys. Condens. Matter* **21**, 395502 (2009).
8. Wang, Y., Lv, J., Zhu, L. & Ma, Y. Crystal structure prediction via particle-swarm optimization. *Phys Rev B* **82**, 094116 (2010).
9. Wang, Y., Lv, J., Zhu, L. & Ma, Y. CALYPSO: A method for crystal structure prediction. *Comput. Phys. Commun.* **183**, 2063–2070 (2012).
10. Zheng, Q. *et al.* Ab initio nonadiabatic molecular dynamics investigations on the excited carriers in condensed matter systems. *WIREs Comput. Mol. Sci.* **9**, e1411 (2019).
11. Huang, -Menglin *et al.* DASP: defect and dopant ab-initio simulation package. - *Journal of Semiconductors* vol. 43 042101 (2022).

12. Wang, V., Xu, N., Liu, J.-C., Tang, G. & Geng, W.-T. VASPKIT: A user-friendly interface facilitating high-throughput computing and analysis using VASP code. *Comput Phys Commun* **267**, 108033 (2021).
13. Hohenberg, P. & Kohn, W. Inhomogeneous electron gas. *Phys. Rev.* **136**, B864–B871 (1964).
14. Kohn, W. & Sham, L. J. Self-consistent equations including exchange and correlation effects. *Phys. Rev.* **140**, A1133–A1138 (1965).
15. Jain, A. *et al.* The Materials Project: a materials genome approach to accelerating materials innovation. *APL Mater.* **1**, 011002 (2013).
16. He, B. *et al.* High-throughput screening platform for solid electrolytes combining hierarchical ion-transport prediction algorithms. *Sci. Data* **7**, 151 (2020).
17. Kim, S. *et al.* A band-gap database for semiconducting inorganic materials calculated with hybrid functional. *Sci. Data* **7**, 387 (2020).
18. de Pablo, J. J., Jones, B., Kovacs, C. L., Ozolins, V. & Ramirez, A. P. The Materials Genome Initiative, the interplay of experiment, theory and computation. *Curr. Opin. Solid State Mater. Sci.* **18**, 99–117 (2014).
19. The materials genome initiative at the national science foundation: a status report after the first year of funded research. *JOM* **66**, 336–344 (2014).
20. Kirklin, S. *et al.* The Open Quantum Materials Database (OQMD): assessing the accuracy of DFT formation energies. *Npj Comput. Mater.* **1**, 15010 (2015).
21. Draxl, C. & Scheffler, M. NOMAD: The FAIR concept for big data-driven materials science. *MRS Bull.* **43**, 676–682 (2018).
22. Agrawal, A. & Choudhary, A. Perspective: materials informatics and big data: realization of the "fourth paradigm" of science in materials science. *APL Mater.* **4**, 053208 (2016).
23. Butler, K. T., Davies, D. W., Cartwright, H., Isayev, O. & Walsh, A. Machine learning for molecular and materials science. *Nature* **559**, 547–555 (2018).
24. Chen, A., Zhang, X. & Zhou, Z. Machine learning: accelerating materials development for energy storage and conversion. *InfoMat* **2**, 553–576 (2020).
25. Han, Y. *et al.* Machine learning accelerates quantum mechanics predictions of molecular crystals. *Phys. Rep.* **934**, 1–71 (2021).
26. Wang, Z., Han, Y., Cai, J., Chen, A. & Li, J. Vision for energy material design: a roadmap for integrated data-driven modeling. *J. Energy Chem.* **71**, 56–62 (2022).
27. Zou, X. *et al.* Machine learning analysis and prediction models of alkaline anion exchange membranes for fuel cells. *Energy Environ. Sci.* **14**, 3965–3975 (2021).
28. Zhang, H., Wang, Z., Ren, J., Liu, J. & Li, J. Ultra-fast and accurate binding energy prediction of shuttle effect-suppressive sulfur hosts for lithium-sulfur batteries using machine learning. *Energy Storage Mater.* **35**, 88–98 (2021).
29. Jiang, B. *et al.* Bayesian learning for rapid prediction of lithium-ion battery-cycling protocols. *Joule* **5**, 3187–3203 (2021).
30. Lyu, R., Moore, C. E., Liu, T., Yu, Y. & Wu, Y. Predictive design model for low-dimensional organic–inorganic halide perovskites assisted by machine learning. *J. Am. Chem. Soc.* **143**, 12766–12776 (2021).
31. Wang, Z., Cai, J., Wang, Q., Wu, S. & Li, J. Unsupervised discovery of thin-film photovoltaic materials from unlabeled data. *Npj Comput. Mater.* **7**, 128 (2021).
32. Miyake, Y. & Saeki, A. Machine learning-assisted development of organic solar cell materials: issues, analyses, and outlooks. *J. Phys. Chem. Lett.* **12**, 12391–12401 (2021).
33. Batra, R., Song, L. & Ramprasad, R. Emerging materials intelligence ecosystems propelled by machine learning. *Nat. Rev. Mater.* **6**, 655–678 (2021).

34. Lu, H. *et al.* Machine learning-aided engineering of hydrolases for PET depolymerization. *Nature* **604**, 662–667 (2022).
35. Wang, G. *et al.* ALKEMIE: An intelligent computational platform for accelerating materials discovery and design. *Comput. Mater. Sci.* **186**, 110064 (2021).
36. Zhao, X.-G. *et al.* JAMIP: an artificial-intelligence aided data-driven infrastructure for computational materials informatics. *Sci. Bull.* **66**, 1973–1985 (2021).
37. Ward, L. *et al.* Matminer: An open source toolkit for materials data mining. *Comput. Mater. Sci.* **152**, 60–69 (2018).
38. Ong, S. P. *et al.* Python materials genomics (pymatgen): a robust, open-source python library for materials analysis. *Comput. Mater. Sci.* **68**, 314–319 (2013).
39. Pedregosa, F. *et al.* Scikit-learn: machine learning in python. *J. Mach. Learn. Res.* **12**, 2825–2830 (2011).
40. Chen, T. & Guestrin, C. XGBoost. *Proc. 22nd ACM SIGKDD Int. Conf. Knowl. Discov. Data Min.* (2016).
41. Mentel, Ł. mendeleev – A Python resource for properties of chemical elements, ions and isotopes. (2014).
42. Meredig, B. *et al.* Combinatorial screening for new materials in unconstrained composition space with machine learning. *Phys Rev B* **89**, 094104 (2014).
43. Wang, Z., Zhang, H. & Li, J. Accelerated discovery of stable spinels in energy systems via machine learning. *Nano Energy* **81**, 105665 (2021).
44. Cai, J., Wang, Z., Wu, S., Han, Y. & Li, J. A machine learning shortcut for screening the spinel structures of Mg/Zn ion battery cathodes with a high conductivity and rapid ion kinetics. *Energy Storage Mater.* **42**, 277–285 (2021).
45. Wang, Z. *et al.* Harnessing artificial intelligence to holistic design and identification for solid electrolytes. *Nano Energy* **89**, 106337 (2021).
46. Zhang, H., Wang, Z., Cai, J., Wu, S. & Li, J. Machine-learning-enabled tricks of the trade for rapid host material discovery in Li–S battery. *ACS Appl. Mater. Interfaces* **13**, 53388–53397 (2021).
47. Zhuo, Y., Mansouri Tehrani, A. & Brgoch, J. Predicting the band gaps of inorganic solids by machine learning. *J. Phys. Chem. Lett.* **9**, 1668–1673 (2018).
48. Petretto, G. *et al.* High-throughput density-functional perturbation theory phonons for inorganic materials. *Sci. Data* **5**, 180065 (2018).
49. Petousis, I. *et al.* High-throughput screening of inorganic compounds for the discovery of novel dielectric and optical materials. *Sci. Data* **4**, 160134 (2017).
50. Zhang, L. *et al.* A database of ionic transport characteristics for over 29 000 inorganic compounds. *Adv. Funct. Mater.* **30**, 2003087 (2020).
51. Zhu, T. *et al.* Charting lattice thermal conductivity for inorganic crystals and discovering rare earth chalcogenides for thermoelectrics. *Energy Environ. Sci.* **14**, 3559–3566 (2021).
52. Yu, J. *et al.* Finding optimal mid-infrared nonlinear optical materials in germanates by first-principles high-throughput screening and experimental verification. *ACS Appl. Mater. Interfaces* **12**, 45023–45035 (2020).
53. Wang, Y., Schwartz, J., Gim, J., Hovden, R. & Mi, Z. Stable unassisted solar water splitting on semiconductor photocathodes protected by multifunctional GaN nanostructures. *ACS Energy Lett.* **4**, 1541–1548 (2019).
54. Zhang, Y. *et al.* Synthesis and characterization of spinel cobaltite ($Co_3O_4$) thin films for function as hole transport materials in organometallic halide perovskite solar cells. *ACS Appl. Energy Mater.* **3**, 3755–3769 (2020).
55. He, X. *et al.* The passivity of lithium electrodes in liquid electrolytes for secondary batteries. *Nat. Rev.*

*Mater.* **6**, 1036–1052 (2021).
56. Wang, Z. *et al.* Computational screening of spinel structure cathodes for Li-ion battery with low expansion and rapid ion kinetics. *Comput. Mater. Sci.* **204**, 111187 (2022).
57. Balaish, M. *et al.* Processing thin but robust electrolytes for solid-state batteries. *Nat. Energy* **6**, 227–239 (2021).
58. Chen, Y.-T. *et al.* Fabrication of high-quality thin solid-state electrolyte films assisted by machine learning. *ACS Energy Lett.* **6**, 1639–1648 (2021).
59. Borlido, P. *et al.* Exchange-correlation functionals for band gaps of solids: benchmark, reparametrization and machine learning. *Npj Comput. Mater.* **6**, 96 (2020).
60. Borlido, P. *et al.* Large-scale benchmark of exchange–correlation functionals for the determination of electronic band gaps of solids. *J. Chem. Theory Comput.* **15**, 5069–5079 (2019).
61. Wang, Z. *et al.* Deep learning for ultra-fast and high precision screening of energy materials. *Energy Storage Mater.* **39**, 45–53 (2021).
62. Sun, H. & Zhao, K. Electronic structure and comparative properties of LiNi$_x$Mn$_y$Co$_z$O$_2$ cathode materials. *J. Phys. Chem. C* **121**, 6002–6010 (2017).
63. Chakraborty, A. *et al.* Layered cathode materials for lithium-ion batteries: review of computational studies on LiNi$_{1-x-y}$Co$_x$Mn$_y$O$_2$ and LiNi$_{1-x-y}$Co$_x$Al$_y$O$_2$. *Chem. Mater.* **32**, 915–952 (2020).
64. Kim, T., Drakopoulos, S. X., Ronca, S. & Minnich, A. J. Origin of high thermal conductivity in disentangled ultra-high molecular weight polyethylene films: ballistic phonons within enlarged crystals. *Nat. Commun.* **13**, 2452 (2022).
65. Zhou, Y., Dong, Z.-Y., Hsieh, W.-P., Goncharov, A. F. & Chen, X.-J. Thermal conductivity of materials under pressure. *Nat. Rev. Phys.* **4**, 319–335 (2022).
66. Zhou, J. *et al.* 2DMatPedia, an open computational database of two-dimensional materials from top-down and bottom-up approaches. *Sci. Data* **6**, 86 (2019).